\documentclass[11pt,a4paper]{article}
\usepackage[hyperref]{acl2020}
\usepackage{times}
\usepackage{latexsym}

\usepackage{amsmath} 
\usepackage[]{algorithm2e}
\usepackage{url}
\usepackage{multirow}
\usepackage{soul}
\usepackage{graphicx}
\usepackage{bm}
\usepackage{balance}

\usepackage{microtype}

\aclfinalcopy 

\title{Review-based Question Generation with Adaptive Instance \\ Transfer and Augmentation\thanks{~~The work described in this paper is partially supported by a grant from the Research Grant Council of the Hong Kong Special Administrative Region, China (Project Code: 14204418).}}
\author{\textbf{Qian Yu\textsuperscript{\rm 1}\thanks{~The work was done when Qian Yu was an intern at Alibaba.}~\ \ \ Lidong Bing\textsuperscript{\rm 2}\ \ \ Qiong Zhang\textsuperscript{\rm 2}\ \ \ Wai Lam\textsuperscript{\rm 1}\ \ \ Luo Si\textsuperscript{\rm 2}}\\
\textsuperscript{\rm 1} The Chinese University of Hong Kong\\
\textsuperscript{\rm 2} DAMO Academy, Alibaba Group\\
\textsuperscript{\rm 1}\texttt{\{yuqian, wlam\}@se.cuhk.edu.hk}\\ \textsuperscript{\rm 2}\texttt{\{l.bing, qz.zhang, luo.si\}@alibaba-inc.com} 
}

\begin{document}

\maketitle
\begin{abstract}
While online reviews of products and services become an important information source, it remains inefficient for potential consumers to exploit verbose reviews for fulfilling their information need. We propose to explore question generation as a new way of review information exploitation, namely generating questions that can be answered by the corresponding review sentences. One major challenge of this generation task is the lack of training data, i.e. explicit mapping relation between the user-posed questions and review sentences. To obtain proper training instances for the generation model, we propose an iterative learning framework with adaptive instance transfer and augmentation.
To generate to the point questions about the major aspects in reviews, related features extracted in an unsupervised manner are incorporated without the burden of aspect annotation.
Experiments on data from various categories of a popular E-commerce site demonstrate the effectiveness of the framework,
as well as the potentials of the proposed review-based question generation task.
\end{abstract}

\section{Introduction}
The user-written reviews for products or service have become an important information source and there are a few research areas analyzing such data, including aspect extraction \cite{bing_TOIT_2016,chen2013exploiting}, product recommendation \cite{Chelliah:2017:PRE:3109859.3109936}, and sentiment analysis \cite{li-etal-2018-transformation,zhao2018weakly}. Reviews reflect certain concerns or experiences of users on products or services, and such information is valuable for other potential consumers. 
However, there are few mechanisms assisting users for efficient review digestion. It is time-consuming for users to locate critical review parts that they care about, particularly in long reviews.

We propose to utilize question generation (QG) \cite{du2017learning} as a new means to overcome this problem.
Specifically, given a review sentence, the generated question is expected to ask about the concerned aspect of this product, from the perspective of the review writer. 
Such question can be regarded as a reading anchor of the review sentence, and it is easier to view and conceive due to its concise form. As an example, the review for a battery case product in Table \ref{tab:example} is too long to find sentences that can answer a user question such as ``How long will the battery last?''. Given the generated questions in the right column, it would be much easier to find out the helpful part of the review. Recently, as a topic attracting significant research attention, question generation is regarded as a dual task of reading comprehension in most works, namely generating a question from a sentence with a fixed text segment in the sentence designated as the answer \cite{duan2017question,sun2018answer}.



\definecolor{capri}{rgb}{0.0, 0.75, 1.0}
\definecolor{darkpink}{rgb}{0.91, 0.33, 0.5}

\begin{table*}[t]
    \centering
    \begin{footnotesize}
    \begin{tabular}{p{320pt}|p{120pt}}\hline
    \hfil Review & \hfil Question\\\hline

   \textcolor{darkpink}{It doesn't heat up like most of the other ones, and I was completely fascinated by the ultra light and sleek design for the case.} \textcolor{capri}{Before I was using the Mophie case but I couldn't wear it often because it was like having a hot brick in your pocket, hence I had to always leave it at home.} {On the contrary, with PowerBear, I never take it off because I can't even tell the difference.} \textcolor{darkpink}{Also it is build in a super STRONG manner and even though I dropped my phone a few times, its shock resistant technology won't let a single thing happen to the case or the phone.} \textcolor{capri}{The PowerBear case became an extension to my phone that I never have to take off because when I charge it at night, it charges both my phone and the case.} \textcolor{darkpink}{I have battery life for more than two days for normal use, i.e. not power-consuming gaming.}
    &  \textcolor{darkpink}{Does this make the phone warm during charging?} \newline \textcolor{capri}{Have any of you that own this had a Mophie?}  \newline \textcolor{darkpink}{Does this give protection to the phone?} \newline \textcolor{capri}{Can this charge the phone and the extra battery at the same time?} \newline \textcolor{darkpink}{How many days it can last?}
     \\\hline
    \end{tabular}\vspace{-2mm}
    \end{footnotesize}
    \caption{A product review and the example questions.}\vspace{-3mm}
    \label{tab:example}
\end{table*}

Two unique characteristics of our review-based question generation task differentiate it from the previous question generation works. 
First, there is no review-question pairs available for training, thus a simple Seq2Seq-based question generation model for learning the mapping from the input (i.e. review) to the output (i.e. question) cannot be applied. Even though we can easily obtain large volumes of user-posed review sets and question sets, they are just separate datasets and cannot provide any supervision of input-output mapping (i.e.  review-question pair).
The second one is that different from the traditional question generation, the generated question from a review sentence will not simply take a fixed text segment in the review as its answer.
The reason is that some reviews describing user experiences are highly context-sensitive. For the example in Table \ref{tab:example}, for the review ``I have battery life for more than two days for normal use, i.e. not power-consuming gaming.'' and its corresponding example question ``How many days it can last?'', obviously the text segment ``more than two days'' is a less precise answer, while the whole review sentence is much more informative. 
In some other case, even such less precise answer span cannot be extracted from the review sentence, e.g. for the example question ``Does this give protection to the phone?'' and the review sentence ``Also it is ... even though I dropped my phone ..., its shock resistant technology won't let a single thing happen to the case or the phone.''. Of course here, a simple ``Yes'' or ``No'' answer does not make much sense as well, while the whole review sentence is a vivid and informative answer.

The above two unique characteristics raise two challenges for our task. The first challenge, namely lacking review-question pairs as training instances, appears to be intractable, particularly given that the current end-to-end models are very data-hungry.
One instant idea is to utilize user-posed (question, answer) pairs as substitute for training. However, several instance-related defects hinder the learned generation model from being competent for the review-based question generation.
Some answers are very short, e.g. ``more than two days'', therefore, without necessary context, they are not helpful to generate good questions. 
The second challenge, namely the issue that some verbose answers contain irrelevant content especially for subjective questions.
To handle this challenge, we propose a learning framework with adaptive instance transfer and augmentation.

Firstly, a pre-trained generation model based on user-posed answer-question pairs is utilized as an initial question \textit{generator}.
A \textit{ranker} is designed to work together with the \textit{generator} to improve the training instance set by distilling it via removing unsuitable answer-question pairs to avoid ``negative transfer'' \cite{pan2009survey}, and augmenting \cite{kobayashi2018contextual} it by adding suitable review-question pairs. 
For selecting suitable reviews for question generation, the \textit{ranker} considers two factors: the major aspects in a review and the review's suitability for question generation. The two factors are captured via a reconstruction objective and a reinforcement objective with reward given by the \textit{generator}. Thus, the \textit{ranker} and the \textit{generator} are iteratively enhanced, and the adaptively transferred answer-question pairs and the augmented review-question pairs gradually relieve the data lacking problem.

In accordance with the second characteristic of our task, it is plausible to regard a review sentence or clause as the answer to the corresponding question originated from it. Such treatment brings in the second challenge: how can we guarantee that the generated question concentrates on the critical aspect mentioned by the review sentence? For example, a question like ``How was the experience for gaming?'' is not a favourable generation for ``I have battery life for more than two days for normal use, i.e. not power-consuming gaming.''. To solve this problem, we incorporate aspect-based feature discovering in the \textit{ranker}, and then we integrate the aspect features and an aspect pointer network in the \textsl{generator}. The incorporation of such aspect-related features and structures helps the \textit{generator} to focus more on critical product aspects, other than the less important parts, which is complied with the real user-posed questions.

To sum up, our main contributions are threefold. (1) A new practical task, namely question generation from reviews without annotated instance, is proposed and it has good potential for multiple applications. (2) A novel adaptive instance transfer and augmentation framework is proposed for handling the data lacking challenge in the task. (3) Review-based question generation is conducted on E-commerce data of various product categories.

\begin{figure*}[ht]
    \centering
    \includegraphics[width=460pt]{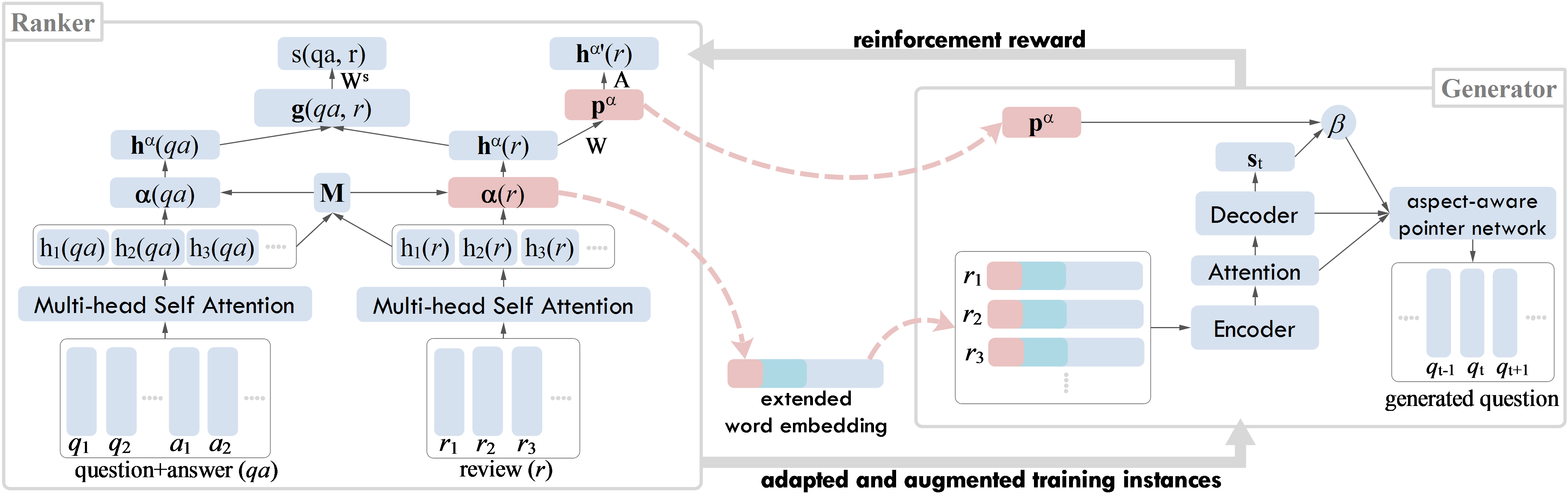}
    \caption{AITA framework. \textbf{M} is the shared parameter matrix for QA and review. 
    \label{fig:framework}}\vspace{-3mm}
\end{figure*}

\section{Related Work}


Question generation (QG) is an emerging research topic due to its wide application scenarios such as education \cite{wang2018qg}, goal-oriented dialogue \cite{lee2018answerer}, and question answering \cite{duan2017question}.
The preliminary neural QG models \cite{du2017learning,zhou2017neural,du2017identifying} outperform the rule-based methods relying on hand-craft features, and thereafter various models have been proposed to further improve the performance via incorporating question type \cite{dong2018neural}, answer position \cite{sun2018answer}, long passage modeling \cite{zhao2018paragraph},  question difficulty \cite{Gao2019DQG}, and to the point context \cite{LiGao2019tothepoint}. Some works try to find the possible answer text spans for facilitating the learning \cite{wang2019a}. Question generation models can be combined with its dual task, i.e., reading comprehension or question answering with various motivations, such as improving auxiliary task performance \cite{duan2017question,yang2017semi,golub2017two}, collaborating QA and QG model \cite{tang2018learning,tang2017question}, and unified learning \cite{xiao2018dual}.

Although question generation has been applied on other datasets, e.g., Wikipedia \cite{du2018harvesting}, most of the existing QG works treat it as a dual task of reading comprehension \cite{yu2018qanet,cui2017attention}, namely generating a question from a piece of text where a certain text span is marked as answer, in spite of several exceptions where only sentences without answer spans are used for generating questions \cite{du2017learning,chali2018automatic}. Such generation setting is not suitable for reviews due to the lack of (question, review) pairs and improper assumption of text span answer as aforementioned.
There are works training the question generation model with the user-written QA pairs in E-commerce sites \cite{hu2018aspect,chali2018automatic}, but the practicality is limited since the questions are only generated from answers instead of reviews.

Transfer learning \cite{pan2009survey,tan2017distant,Juntao2020CDCL} refers to a broad scope of methods that exploit knowledge across domains for handling tasks in the target domain. A few terms are used for describing specific methods in this learning paradigm, e.g., self-taught learning \cite{raina2007self}, domain adaptation \cite{long2017deep}, etc. Based on ``what to transfer'', transfer learning is categorized into four groups \cite{pan2009survey}, namely instance transfer, feature representation transfer, parameter transfer, and relational knowledge transfer. Our learning framework can be regarded as a case of instance transfer with iterative instance adaptation and augmentation.


\section{The Proposed AITA Framework}


For handling the aforementioned issues, we propose an Adaptive Instance Transfer and Augmentation (AITA) framework as shown in Figure~\ref{fig:framework}. Since the review-related processing is always sentence-based, we use ``review'' for short to refer to review sentence in this paper.
Its two components, namely \textit{ranker} and \textit{generator}, are learned iteratively.
Initially, AITA simply transfers all available (question, answer) pairs and trains a \textit{generator}. 
Then it will iteratively enhance the \textit{generator} with the help of the \textit{ranker}.
The \textit{ranker} takes a (question, answer) pair and a review as its input and calculates a ranking score $s$. Thus, it can rank all reviews for a given QA pair. 
The ranking objective incorporates the reward provided by the \textit{generator}, which helps find out those suitable reviews to form (review, question) pairs for training (i.e. augmenting the training data).
Meanwhile, the reward from the generator also helps remove unsuitable QA pairs for training, so that it makes the transfer more adaptive. 
Note that the \textit{ranker} also learns to model two hidden aspect related variables for the review, which are helpful for the \textit{generator} to ask about the major aspects in review. Such an iterative instance manipulation procedure gradually transfers and augments the training set for handling review-based question generation.

\subsection{Review Ranker for Data Augmentation}
\label{sec:sec:ranker}
There are two pieces of input text for \textit{ranker}. The first one is the concatenation of a (question, answer) pair $qa$ and the second one is a review sentence $r$. $qa$ and $r$ are associated with the same product.
Since the \textit{ranker} is responsible for instance augmentation that provides (question, review) pairs, it is trained to learn a score $s(qa, r)$ which can be used to return suitable $r$'s for a given $qa$.

\paragraph{Ranking with Partially Shared Encoders.}
The input $qa$ and $r$ are encoded with two Transformer encoders with the same structure and partially shared parameters, to leverage the advantage of multi-head self attention on modeling word associations without considering term position.
An input ($qa$ or $r$) is written as a matrix $\text{E} = [e_1^T, ..., e_n^T]^T$, where $e$ is a word embedding and $n$ is the text length. The number of heads in the multi-head self-attention is denoted as $m$, and the output of the $j$-th head is written as:
\begin{align}
\text{Q}^j, \text{K}^j, \text{V}^j &= \text{E}\text{W}^j_Q, \text{E}\text{W}^j_K, \text{E}\text{W}^j_V\\
\text{head}^j(\text{E}) &= \text{softmax}(\frac{\text{Q}^j\text{K}^{jT}}{\sqrt{d}})\text{V}^j
\end{align}
where $d$ is the dimension of word embedding.
The outputs of different heads are concatenated and the encoding for the $i$-th word is written as $\textbf{h}_i = [\text{head}^1_i; ...; \text{head}^m_i]$.

To obtain the sentence representation considering the complete semantics, we apply a global attention layer on the output of the Transformer encoder:
\begin{equation}
    \textbf{h}^\alpha = \sum_{i=1}^n \alpha_i \textbf{h}_i
    \label{eqn:encoding}
\end{equation}
where the attention weight $\alpha_i = \exp(\textbf{h}_i\cdot \text{M}\cdot \overline{\textbf{h}})/Z_\alpha$, $Z_\alpha$ is the normalization, and $\overline{\textbf{h}} = \sum{\textbf{h}_i}/n$. The parameter matrix $\text{M}$ is shared by encoders for both $qa$ and $r$ for capturing the common attention features across them.

After encoding $qa$ and $r$ as $\textbf{h}^\alpha(qa)$ and $\textbf{h}^\alpha(qa)$, a vector $\textbf{g}(qa, r)$ is assigned with the concatenation of $\textbf{h}^\alpha(qa)$, $\textbf{h}^\alpha(qa)$ and their difference
\begin{equation}
    \textbf{g}(qa,r) = \left[\ \textbf{h}^\alpha(qa), \textbf{h}^\alpha(r), |\textbf{h}^\alpha(qa)-\textbf{h}^\alpha(r)| \ \right]\notag
\end{equation}
The review ranking score $s(qa, r)$ is calculated as:
\begin{gather}
    s(qa, r) = \sigma(\text{W}^s\textbf{g}(qa, r) + \textbf{b}^s)
\end{gather}
where $\sigma$ is sigmoid function.

\paragraph{Reinforcement Objective for Ranker Learning.}
To learn an appropriate $s(qa, r)$, we encounter a major challenge, namely lacking ground truth labels for (question, review). Our solution takes the \textit{generator} in our framework as an agent that can provide reward for guiding the learning of \textit{ranker}. The \textit{generator} is initially trained with (question, answer) data, and is gradually updated with adapted and augmented training instances, so that the rewards from the \textit{generator} can reflect the ability of review for generating the corresponding question. 

Specifically, we propose a reinforcement objective that makes use of the reward from the \textit{generator}, denoted as $\text{reward}_G(r,q)$.
For each pair of question and review, we take the normalized $\log \text{ppl}(q|r)$ in the \textit{generator} as reward:
\begin{equation}
    \text{reward}_G(r,q) = \frac{\log \text{ppl}(q|r)}{\sum_{r^*\in R_{qa}}\log \text{ppl}(q|r^*)}
\end{equation}
where $\text{R}_{qa}$ is the reviews under the same product as $qa$, and $\log \text{ppl}(q|r)$ is the log perplexity of generating a question $q$ from a review $r$:
\begin{equation}
        \log \text{ppl}(q|r)
    = -\frac{1}{|q|}\sum_{t\in [1,|q|]}p_G(q_t|r,q_1...q_{t-1})\nonumber
\end{equation}

The reinforcement objective for the \textit{ranker} is to maximize the average reward for all the reviews given a question. The sampling probabilities for reviews are obtained via normalized ranking score, namely $p(r|qa) = s(qa, r) / \text{Z}_{qa}$, where $\text{Z}_{qa} = \sum_{r* \in \text{R}_{qa}} s(qa, r*)$.
The loss function is:
\begin{equation}
    L^{g}(qa, r) = \text{E}_{r\sim p(r|qa)}\text{reward}_G(r,q)
\end{equation}
The gradient calculation for the above objective is an intractable problem. As an approximated method which performs well in the iterative algorithm, the normalization term $\text{Z}_{qa}$ is fixed during the calculation of the policy gradient:
\begin{equation}
    \Delta L^{g}(qa, r) = \sum_r \Delta s(qa, r) \text{reward}_G(r,q) / \text{Z}_{qa}\notag
\end{equation}

\paragraph{Regularization with Unsupervised Aspect Extraction.} 
Product aspects usually play a major role in all of product questions, answers and reviews, since they are the discussion focus of such text content. Thus, such aspects can act as connections in modeling input pairs of $qa$ and $r$ via the partially shared structure. 
To help the semantic vector $\textbf{h}^\alpha$ in Eqn \ref{eqn:encoding} capture salient aspects of reviews, an autoencoder module is connected to the encoding layer for reconstructing $\textbf{h}^{\alpha}$. Together with the matrix $\text{M}$, the autoencoder can be used to extract salient aspects from reviews. Note that this combined structure is similar to the ABAE model \cite{he2017unsupervised}, which has been shown effective for unsupervised aspect extraction. Compared with supervised aspect detection methods, such a unsupervised module avoid the burden of aspect annotation for different product categories, and our experiments demonstrate that regularization based on this module is effective.

Specifically, $\textbf{h}^\alpha$ is mapped to an aspect distribution $\textbf{p}^\alpha$ and then reconstructed:
\begin{gather}
    \textbf{p}^\alpha = \text{softmax}(\text{W}^p\cdot \textbf{h}^\alpha + \textbf{b}^p)\\\label{eqn:dist}
    {\textbf{h}^{\alpha}}' = \textbf{p}^\alpha \cdot \text{A}
\end{gather}
where each dimension in $\bm{p}^\alpha$ stands for the probability that the review contains the corresponding aspect, and ${\bm{h}^\alpha}'$ is the reconstruction of review representation, and $\text{A}$ is a learnable parameter matrix. Note that we define ``aspects'' as implicit aspect categories, namely clusters of associated attributes of product, which is commonly used in unsupervised aspect extraction  \cite{wang2015sentiment,he2017unsupervised}.
The reconstruction objective is written as:
\begin{equation}
    L^{\alpha}(qa, r) = [\textbf{h}^\alpha(r) - {\textbf{h}^\alpha}'(r)]^2 \  / \ 2.
\end{equation}
Only the reconstruction of review representations is considered since we focus on discovering aspects in reviews.\footnote{We simplified the objective in AEAB model by eliminating the additional regularization term which is not necessary when combining $L^{\alpha}(qa, r)$ and $L^{g}(qa, r)$.}
In this way, the aspect-based reconstruction will force $\textbf{h}^\alpha$ to focus on salient aspects that facilitate the reconstruction.
The final loss function of the \textit{ranker} is regularized to:
\begin{equation}\label{eqn:obj}
    L(qa, r) = L^{g}(qa, r) - \lambda L^{\alpha}(qa, r)
\end{equation}
where $\lambda$ is a hyper-parameter.




\subsection{Question Generator in Transfer Learning}
\label{sec:sec:generator}
We adapt the Seq2Seq model for the aspect-focused generation model, which is updated gradually via the transferred and augmented instances. With the help of aspect-based variables learned in \textit{ranker}, the \textit{generator} can generate questions reflecting the major aspect in the review.


\paragraph{Aspect-enhanced Encoding.} To emphasize the words related to salient aspects, the attention weight $\alpha_i$ obtained in the \textit{ranker} is incorporated into the word embedding. Given an input review sentence, we obtain the extended word embedding $\tilde{\textbf{e}}_i$ at position $i$:
\begin{equation}
    \tilde{\textbf{e}}_i = [\textbf{e}_i, \textbf{e}_i^{POS}, \textbf{e}_i^{NER}, \alpha_i]
\end{equation}
where $\textbf{e}_i$ is the pre-trained word embedding, $\textbf{e}_i^{POS}$ is the one-hot POS tag of $i$-th word, $\textbf{e}_i^{NER}$ is a BIO feature for indicating whether the $i$-th word is a named entity, and $\alpha_i$ indicates the aspect-based weight for the $i$-th word.
Bi-LSTM is adopted as the basic encoder of \textit{generator}, encoding the $i$-th word as the concatenation of hidden states with both directions: $\textbf{h}_i^g = [\overrightarrow{\textbf{h}}_i, \overleftarrow{\textbf{h}}_i]$.


\paragraph{Decoding with Aspect-aware Pointer Network.}
Pointer network, i.e., copy mechanism, can significantly improve the performance of text generation. In our task, in addition to the word-level hidden state in the decoder, the overall aspect distribution of the review can also provide clues for how likely the \textit{generator} should copy corresponding review aspect words into the generated question.

The question is generated with an LSTM decoder. The word probability for the current time step is formulated as:
\begin{equation}
    p_0(q_t) = \text{softmax}(\text{W}_2 \tau + \textbf{b}_2)\notag
\end{equation}
and related variables are calculated as:
\begin{gather}
\small
    \tau = \sigma(\text{W}_1[\textbf{s}_t, \textbf{c}_t]+\textbf{b}_1)\ ,\ \ \ \ \ 
   \textbf{s}_t = \text{LSTM}(y_t, \textbf{s}_{t-1})\ ,\notag\\ 
   \textbf{c}_t = \sum_j\textbf{z}_{tj}\textbf{h}^g_j\ , \ \ \ \ \ \ \ \ \ \ 
   \textbf{z}_{tj} = \text{softmax}(\textbf{h}^g_j\text{W}_h\textbf{s}_t) \notag
\end{gather}
where $\textbf{s}_t$ is the hidden state for the $t$-th word in question and $\textbf{c}_t$ is the context encoding based on attention weight $\textbf{z}_{tj}$.

In the pointer network, for a particular position $t$ in the generated text, the word may be copied from a distribution based on the attention weight $\textbf{z}_t$=$\{z_{tj}\}$, where the copy probability is assigned according to the current hidden state $\textbf{s}_t$. We also consider the influence of the aspect distribution $\textbf{p}^\alpha$ in the copy probability $\beta$ for interpolation:
\begin{equation}
    \beta = \sigma(\textbf{p}^\alpha\text{W}_c\textbf{s}_t + \textbf{b}_c)
\end{equation}
The incorporation of $\textbf{p}^\alpha$ helps the pointer network to consider the overall aspect distribution of context in addition to the semantics in the current position for copying words. Finally, the $t$-th word is generated from the mixture of the two distributions:
\begin{equation}
    p(q_t) = (1-\beta)\cdot p_0(q_t) + \beta\cdot \textbf{z}_{t}.
\end{equation}

The \textit{generator} is trained via maximizing the likelihood of the question $q$ given the review $r$:
\begin{equation}
p(r|q) = \sum_i p(r_i | q, r_1, ..., r_{i-1})
\end{equation}

\SetEndCharOfAlgoLine{}
\SetKwRepeat{Do}{Repeat}{Until}%
\SetKwFor{For}{For}{Do}{End}%
\SetKwFor{Stageone}{First Stage:}{}{End First Stage}
\SetKwFor{Stagesec}{Second Stage:}{}{End Second Stage}
\begin{algorithm}[!t]


 \KwData{QA\ set $\textbf{S}_{qa}$=\{($q$,$a$)\}; review set $\textbf{S}_{r}$=\{$r$\}; $\mu$}
 \KwResult{\textbf{S}; \textit{generator} trained with \textbf{S}}
 
 Prepare pairs of {($qa$, $r$)} under each product\;
 Initialize the training set \textbf{S} = $\textbf{S}_{qa}$\;

 \For{\text{each epoch}}{
 1. Train \textit{generator} with \textbf{S}.\;
 2. Prepare the $\text{reward}_G(qa,r)$ as \textit{generator} reward for each pair of ($qa$, $r$) (each answer $a$ in $qa$ pairs is regarded as a review for $q$).\;
 3. Adapt \textbf{S} via removing $\mu$ instances with low reward.\;
 4. Train \textit{ranker} according to the objective in Eqn \ref{eqn:obj}.\;
 5. Augment \textbf{S} via adding $\mu$ pairs of instances, which are ($q$, $r$) pairs with top $s(qa, r)$ in \textit{ranker}.\;
 6. Collect $\bm{\alpha}$ and $\bm{p}^\alpha$ for instances in \textbf{S}\ from \textit{ranker}.
 }
 \caption{Learning algorithm of AITA.\label{alg}}
\end{algorithm}

\subsection{Iterative Learning Algorithm}
\label{sec:sec:algorithm}
The purpose of our iterative learning, as by Alg~\ref{alg}, is to update the \textit{generator} gradually via the instance augmentation. The input data for the iterative learning consists of the transferred instance set of question-answer pairs $\textbf{S}_{qa}$, an unlabeled review set $\textbf{S}_{r}$, and an adaption parameter $\mu$. When the learning is finished, two outputs are produced: the final training instances $\textbf{S}$, and the learned \textit{generator}.
The training set $\textbf{S}$ for \textit{generator} is initialized with $\textbf{S}_{qa}$. In each iteration of the algorithm, the \textit{generator} is trained with current $\textbf{S}$, and then $\textbf{S}$ is adapted accordingly. The \textit{ranker} is trained based on the rewards from the generation, which is used for instance augmentation in $\textbf{S}$. Thus, the training set $\textbf{S}$ is updated during the iterative learning, starting from a pure (question, answer) set. Analysis on the influence of the composition of $\textbf{S}$, i.e., instance numbers of two types, is presented in Section \ref{sec:sec:analysis}.

There are two kinds of updates for the instance set $\textbf{S}$: (1) adaption via removing ($q$, $a$) pairs with low \textit{generator} reward, in order to avoid ``negative transfer''; (2) augmentation via adding ($q$, $r$) pairs that are top ranked by \textit{ranker}, in order to increase the proportion of suitable review-question instances in training set. The instance number hyperparameter $\mu$ for removing and adding can be set according to the scale of $\textbf{S}_{qa}$, and more details are given in our experimental setting.

To guarantee the effective instance manipulation, two interactions exist between \textit{generator} and \textit{ranker}. First, aspect-related variables for reviews obtained by \textit{ranker} are part of the \textit{generator} input. The second interaction is that a reward from \textit{generator} is part of the learning objective for \textit{ranker}, in order to teach \textit{ranker} to capture the suitable reviews for generating the corresponding question.

\section{Experiments}

\subsection{Datasets}

We exploit the user-written QA dataset collected in \cite{wan2016modeling} and the review set collected in \cite{mcauley2015image} as our experimental data. The two datasets are collected from $Amazon.com$ separately. We filter and merge the two datasets to obtain products whose associated QA pairs and reviews can both be found. The statistics for our datasets can be found in Table~\ref{dataset}, where the numbers of product for several very large product categories are restricted to 5000.
According to the average lengths, we can find that the whole review tend to be very long. It justified our assumption that it is not easy for users to exploit reviews, and questions with short length can be a good catalogue for viewing reviews.

To test our question generation framework, we manually labeled 100 ground truth review-question pairs for each product category. 6 volunteers are asked to select user-posed questions and the corresponding review sentences that can serve as answers. Specifically, the volunteers are given pairs of question and review, and only consider the relevance between question and review. The answer to the question is also accessible but it is only used for helping annotators to understand the question. All labeled pairs are validated by two experienced annotators with good understanding for the consumer information need in E-commerce.

. 

The labeled instances are removed from the training set.
\begin{table}[t]
    \centering
    \begin{small}
    \begin{tabular}{|c|c|c|c|c|c|}
    \hline
& \#p    &  \#q      &\#a     & \#r     &\#(s) \\\hline
Auto & 0.8k    & 5.5k   & 18.7k &  9.4k  &  46.5k\ \\
Baby&  1.9k  &  11.9k   &  38.7k   & 75.3k  & 450.7k\\
Beauty & 2.5k  &  15.9k & 53.7k&  62.4k  &  338.6k\\
Phones & 3.6k  & 23.8k  &  87.4k  &   104.5k   &  561.8k\\
Cloth & 0.4k   &  0.30k &   10.7k  &  6.9k  &  32.2k\\
Elec & 5k  &  31.0k   & 101.2k &  229.4k  & 1461.8k\\
Health  &5k  &  32.4k &   114.2k &  136.9k & 749.9k\\
Music & 0.4k  &   2.7k   &  8.9k  & 5.2k   &  27.9k\\
Sports & 5k  &  34.2k  &  120.6k & 122.6k &   648.5k\\
Tools & 4.1k  &  29.8k  &  104.1k &  70.7k &   425.6k\\
\hline
    \end{tabular}

\begin{tabular}{|c|c|c|c|c|}
    \hline
&  $L_q$      & $L_a$ &   $L_{r}$   &  $L_{s}$\\\hline
Auto & \ 14.4\ \   & \ 23.3\ \  & 88.3  & 17.8\ \\
Baby&   15.2  &   22.9   & 106.4   & 17.8 \\
Beauty &  13.1  &   22.0  & 88.6   & 16.3\\
Phones  & 13.2   & 19.2 &   97.0   & 18.1\\
Cloth  &  13.0    &19.8 &  71.2  &  15.3\\
Elec &  16.1   & 24.8  &  119.5  &18.8\\
Health  & 13.0 &  22.5 &  96.0 &   17.5\\
Music  & 14.6  &    24.0 &  94.2   & 17.7\\
Sports &  13.6  & 22.3  & 91.0  & 17.2\\
Tools  &  14.7   &23.2 &  110.2 &  18.3\\
\hline
    \end{tabular}
    
    \begin{tabular}{|c|c|c|c|c|}
    \hline

\hline
    \end{tabular}
    
    \end{small}
    \caption{Data statistics. \#: number; $p$, $q$, $a$, $r$: product, question, answer, whole review; $s$: review sentence, $L_q$, $L_a$, $L_r$, $L_{s}$ are their average lengths.}
    \label{dataset}
\end{table}
\subsection{Experimental Settings}
For each product category, we train the AITA framework and use the learned \textit{generator} for testing. The fixed 300 dimension GloVe word embeddings \cite{Pennington2014GloveGV} are used as the basic word vectors. For all text including question, answer and review, we utilize StanfordNLP for tokenizing, lower casing, and linguistic features extraction, e.g., NER \& POS for the encoder in \textit{generator}. 
In \textit{ranker}, the dimension of aspect distribution is set to 20 and the $\lambda$ in the final loss function in Eqn~\ref{eqn:obj} is set to 0.8. In the multi-head self-attention, the head number is set to 3 and the dimension for Q, K, V is 300. The dimensions of matrices can be set accordingly.
The hidden dimension in \textit{generator} is set to 200. In the iterative learning algorithm, we set the epoch number to 10 and the updating instance number $\mu$ to $0.05 \times |\mathbf{S}_{qa}|$. In testing, given a review $r$ as input for \textit{generator}, the additional input variables $\bm{\alpha}(r)$ and $\textbf{p}^\alpha(r)$ are obtained via the review encoder (Eqn~\ref{eqn:encoding}) and aspect extraction (Eqn~\ref{eqn:dist}), which are question-independent.

For testing the effectiveness of our learning framework and the incorporation of aspect, we compare our method with the following models:
\textbf{G}$_a$ \cite{du2017learning}: A sentence-based Seq2Seq generation model trained with user-written answer-question pairs.
\textbf{G}$_{a}^{PN}$ \cite{wang2018qg}: A pointer network is incorporated in the Seq2Seq decoding to decide whether to copy word from the context or select from vocabulary.
\textbf{G}$_{ar}^{PN}$: Review data is incorporated via a retrieval-based method. Specifically, the most relevant review sentence for each question is retrieved via BM25 method, and such review-question pairs are added into the training set.
\textbf{G}$_{a}^{PN}$+aspect \cite{hu2018aspect}: Aspect is exploited in this model. We trained the aspect module in our framework, i.e. only using the reconstruction objective to obtain an aspect feature extractor from reviews. Then the aspect features and distributions can be used in the same way as in our method.
\textbf{AITA} refers to our proposed framework. \textbf{AITA}-aspect: All the extracted aspect-related features are removed from \textbf{AITA} as an ablation for evaluating the effectiveness of the unsupervised module for aspect. For every product category, we run each model for 3 times and report the average performance with four evaluation metrics, including BLEU1 (B1), BLEU4 (B4), METEOR (MET) and ROUGE-L (R$_L$).

\begin{table*}[t]
    \centering
    \begin{small}
    \begin{tabular}{|c|c|c|c|c|c|c|c|c|}
    \hline
         &{BLEU1}&{BLEU4}&{METEOR}&{ROUGE-L}&{BLEU1}&{BLEU4}&{METEOR}&{ROUGE-L}\\\hline\hline
        & \multicolumn{4}{c|}{Automative} &\multicolumn{4}{c|}{Baby} \\\hline
       G$_a$  & \ 0.103\ \ \ &\ 0.047\ \ \ &\ 0.062\ \ \ &\ 0.089\ \ \ &\ 0.104\ \ \ &\ 0.055\ \ \ &\  0.065\ \ \ &\ 0.068\ \ \ \\
       G$_{a}^{PN}$ & 0.162& 0.090& 0.091& 0.140& 0.153& 0.088& 0.087& 0.195\\
       G$_{ar}^{PN}$ & 0.147& 0.082& 0.078& 0.118& 0.133& 0.060& 0.068& 0.102\\
       G$_{a}^{PN}$+aspect & 0.165& 0.090& 0.093& 0.140& 0.157& 0.088& 0.091& 0.203\\ \hline
       AITA-aspect & {0.179}& {0.094}& {0.094}& {0.146}& {0.157}& \textbf{0.089}& {0.092}& {214}\\
       AITA & \textbf{0.184}& \textbf{0.097}& \textbf{0.099}& \textbf{0.148}& \textbf{0.167}& \textbf{0.089}& \textbf{0.094}& \textbf{0.221}\\
       \hline\hline
         & \multicolumn{4}{c|}{Beauty} &\multicolumn{4}{c|}{Cell Phone} \\\hline
       G$_a$ & 0.133& 0.088& 0.118& 0.218& 0.203& 0.125& 0.130& 0.104\\
       G$_{a}^{PN}$ & 0.235& 0.122& 0.128& 0.257& 0.250& 0.122& 0.150& 0.217\\
       G$_{ar}^{PN}$ & 0.194& 0.098& 0.119& 0.205& 0.215& 0.117& 0.136& 0.141\\
       G$_{a}^{PN}$+aspect & 0.240& 0.122& 0.132& 0.257& 0.251& 0.134& 0.154& 0.223\\ \hline
       AITA-aspect & {0.240}& {0.127}& {0.132}& {0.257}& {0.261}& {0.139}& {0.184}& {0.230}\\
       AITA & \textbf{0.249}& \textbf{0.129}& \textbf{0.136}& \textbf{0.259}& \textbf{0.267}& \textbf{0.142}& \textbf{0.193}& \textbf{0.244}\\
       \hline\hline
         & \multicolumn{4}{c|}{Clothing \& Jewelry} &\multicolumn{4}{c|}{Electronics} \\\hline
       G$_a$  & 0.224& 0.093& 0.091& 0.178& 0.099& 0.048& 0.107& 0.144\\
       G$_{a}^{PN}$ & 0.283& 0.134& 0.118& 0.227& 0.124& 0.069& \textbf{0.131}& 0.171\\
       G$_{ar}^{PN}$ & 0.258& 0.110& 0.101& 0.198& 0.100& 0.053& 0.121& 0.156\\
       G$_{a}^{PN}$+aspect & 0.298& 0.139& 0.125& 0.241& 0.120& 0.069& 0.126& 0.171\\\hline
       AITA-aspect & {0.306}& {0.152}& {0.138}& {0.246}& {0.125}& {0.069}& \textbf{0.131}& {0.174}\\
       AITA & \textbf{0.316}& \textbf{0.157}& \textbf{0.145}& \textbf{0.263}& \textbf{0.127}& \textbf{0.073}& \textbf{0.131}& \textbf{0.175}\\
       \hline\hline
         & \multicolumn{4}{c|}{Health} &\multicolumn{4}{c|}{Musical Instruments} \\\hline
       G$_a$  & 0.114& 0.062& 0.091& 0.095& 0.088& 0.054& 0.096& 0.091\\
       G$_{a}^{PN}$ & 0.130& 0.080& 0.089& 0.108& 0.114& 0.110& 0.121& 0.119\\
       G$_{ar}^{PN}$ & 0.124& 0.069& 0.086& 0.104& 0.090& 0.072& 0.106& 0.103\\
       G$_{a}^{PN}$+aspect & 0.133& 0.100& 0.123& 0.175& 0.118& 0.110& 0.130& 0.192\\\hline
       AITA-aspect & {0.137}& {0.100}& {0.121}& {0.179}& {0.124}& {0.110}& {0.136}& {0.201}\\
       AITA & \textbf{0.142}& \textbf{0.109}& \textbf{0.132}& \textbf{0.194}& \textbf{0.129}& \textbf{0.112}& \textbf{0.141}& \textbf{0.205}\\
       \hline\hline
         & \multicolumn{4}{c|}{Sports \& Outdoors} &\multicolumn{4}{c|}{Tools} \\\hline
       G$_a$  & 0.079& 0.046& 0.042& 0.064& 0.098& 0.059& 0.093& 0.105\\
       G$_{a}^{PN}$ & 0.091& 0.052& 0.079& \textbf{0.102}& 0.107& 0.077& 0.112& 0.135\\
       G$_{ar}^{PN}$ & 0.087& 0.050& 0.071& 0.083& 0.100& 0.072& 0.103& 0.119\\
       G$_{a}^{PN}$+aspect & 0.091& 0.052& 0.079& \textbf{0.102}& 0.110& 0.079& 0.110& 0.136\\\hline
       AITA-aspect & {0.094}& {0.052}& {0.080}& {0.102}& {0.112}& {0.079}& {0.116}& {0.142}\\
       AITA & \textbf{0.097}& \textbf{0.057}& \textbf{0.083}& \textbf{0.102}& \textbf{0.117}& \textbf{0.083}& \textbf{0.120}& \textbf{0.149}\\
       \hline
    \end{tabular}
    \end{small}
    \caption{Overall performance on question generation.}
    \label{tab:result}
\end{table*}

\subsection{Evaluation of Question Generation}
The results are demonstrated in Table~\ref{tab:result}. \textbf{AITA} achieves the best performance on all product categories regarding different evaluation metrics. The significant improvements over other models demonstrate that our instance transfer and augmentation method can indeed reduce inappropriate answer-question pairs and provide helpful review-question pairs for the \textit{generator}. The performance of \textbf{G}$_{a}$ is very poor due to the missing of attention mechanism. Both \textbf{G}$_{a}^{PN}$ and \textbf{G}$_{a}^{PN}$+aspect have worse performance than ours, even though some product categories have large volume of QA pairs ($>$100k), e.g., Electronics, Tools, etc. This indicates that the answer-question instances are not capable of learning a review-based question generator because of the different characteristics between the answer set and review set. \textbf{G}$_{ar}^{PN}$ performs much worse than \textbf{G}$_{a}^{PN}$, which proves that a simple retrieval method is not effective for merging the instances related to reviews and answers. \textbf{AITA} adapts and augments the QA set to select suitable review-question pairs considering both aspect and generation suitability, resulting in a better \textit{generator}. In addition, effectiveness of aspect feature and aspect pointer network can be illustrated via the slight but stable improvement of \textbf{G}$_{a}^{PN}$+aspect over \textbf{G}$_{a}^{PN}$ and the performance drop of \textbf{AITA}-aspect on all the categories. This proves that even without precise aspect annotation, our unsupervised aspect-based regularization is helpful for improving generation.


\begin{table}[!t]
    \centering
    \begin{tabular}{|c|c|c|c|}
    \hline
    \multicolumn{4}{|c|}{{Clothing \& Jewelry}}\\\hline
   & {\textsl{Relevance}}&{\textsl{Aspect}}& {\textsl{Fluency}}\\\hline
   G$_{a}^{PN}$ & 0.58& 0.62 &{2.58} \\
   \ G$_{ar}^{PN}$ & 0.47& 0.58 &{2.29} \\
      \ G$_{a}^{PN}$+aspect & 0.66& 0.72 &{2.76} \\
      
      AITA  & \textbf{0.80}&\textbf{0.80}&\textbf{2.86}\\
    \hline
    \end{tabular}
    \begin{tabular}{|c|c|c|c|}
    \hline
    \multicolumn{4}{|c|}{{Cell Phone}}\\\hline
   &{\textsl{Relevance}}&{\textsl{Aspect}} & {\textsl{Fluency}} \\\hline
   G$_{a}^{PN}$  &0.42&0.55 &2.79\\
   \ G$_{ar}^{PN}$ &0.35&0.41 &2.44\\
      \ G$_{a}^{PN}$+aspect &0.58&0.63 &2.83\\
      
      AITA  &\textbf{0.72}&\textbf{0.72}&\textbf{2.90}\\
    \hline
    \end{tabular}
    \caption{Performance of human evaluation.}
    \label{tab:manual}
\end{table}

\begin{table}[t]
    \centering
    \begin{small}
    \begin{tabular}{@{}p{220pt}@{}}\hline

    \textcolor{capri}{The entire length of the watch is 9 inches, but the effective length from the last hole to clasp is about 8 inches.} \\ 
    \ \ \ \ - \textbf{G}$_{a}^{PN}$: {What is the difference between gear 2 neo and this watch?}\\
    \ \ \ \ - \textbf{G}$_{a}^{PN}$+aspect: {How is the length?}\\
    \ \ \ \ - \textbf{AITA}: {What is the dimension in mm?}\\\hline
    
    \textcolor{capri}{If you have a huge wrist this watch mayn't look good nor fit you well.} \\
    \ \ \ \ - \textbf{G}$_{a}^{PN}$: {What is the wrist size?}\\
    \ \ \ \ - \textbf{G}$_{a}^{PN}$+aspect: {How does it fit?}\\
    \ \ \ \ - \textbf{AITA}: {Will it fit my huge hand?}\\\hline
    
    \textcolor{capri}{The stainless steel case back can be pried off from the 12 o'clock position (from the back), and the battery CAN be replaced.} \\\ \ \ \ - \textbf{G}$_{a}^{PN}$: {Is the material good quality and not easy to tore?}\\
    \ \ \ \ - \textbf{G}$_{a}^{PN}$+aspect: {Can the lid be removed?}\\
    \ \ \ \ - \textbf{AITA}: {Can you tell me how to replace the battery?}\\\hline
    
    \textcolor{capri}{The watch has a Japanese Miyota movement inside, and has a Japanese Sony 626sw battery which requires you to loosen a very small flat head screw and slide a little metal arm out of the way to remove the battery.} \\\ \ \ \ - \textbf{G}$_{a}^{PN}$: {What is the battery life on this watch?}\\
    \ \ \ \ - \textbf{G}$_{a}^{PN}$+aspect: {Can I remove the battery?}\\
    \ \ \ \ - \textbf{AITA}: {Can I remove the battery?}\\\hline
    \end{tabular}
    \end{small}
    \caption{Case study of generated questions.} 
    \label{tab:case}
\end{table}

\subsection{Human Evaluation and Case Study}
We conduct human evaluation on two product categories to study the quality of the generated questions. Two binary metrics \textsl{Relevance} and \textsl{Aspect} are used to indicate whether a question can be answered by the review and whether they share the same or related product aspect. The third metric, \textsl{Fluency} with the value set \{1, 2, 3\}, is adopted for judging the question fluency. 1 means not fluent and 3 means very fluent. We selected 50 generated questions from each model and asked 4 volunteers for evaluation. The average scores are reported in Table \ref{tab:manual}, which shows that our framework achieves the best performance regarding all the metrics, especially for \textsl{Relevance}, showing that our AITA can help generate more accurate questions based on reviews and thus facilitates exploiting reviews. Due to the incorporation of implicit aspect information, both \textbf{AITA} and \textbf{G}$_{a}^{PN}$+aspect significantly outperform \textbf{G}$_{a}^{PN}$ regarding both \textsl{Aspect} and \textsl{Relevance}. 
Again, \textbf{G}$_{ar}^{PN}$ with a simple retrieval method for augmenting training instances cannot perform well.


The blue sentences in Table~\ref{tab:case} are from a long review talking about some important information of a wat
ch, and the questions generated by different models are also given. These questions are more user-friendly and potential consumers can browse them to quickly locate the information they care about. For example, if a user wants to know more about the battery replacement, the portion before the third sentence can be skipped. According to the generated questions via three methods in the Table \ref{tab:case}, we can find that the questions from \text{AITA} are asking about major aspects of the review sentences. \textbf{G}$_{a}^{PN}$ failed to capture major aspects in the first three sentences, and the questions generated by \textbf{G}$_{a}^{PN}$+aspect are not as concrete as ours, owning to the insufficient training instances.


\begin{figure}[!t]
    \centering
    \includegraphics[width=0.38\textwidth]{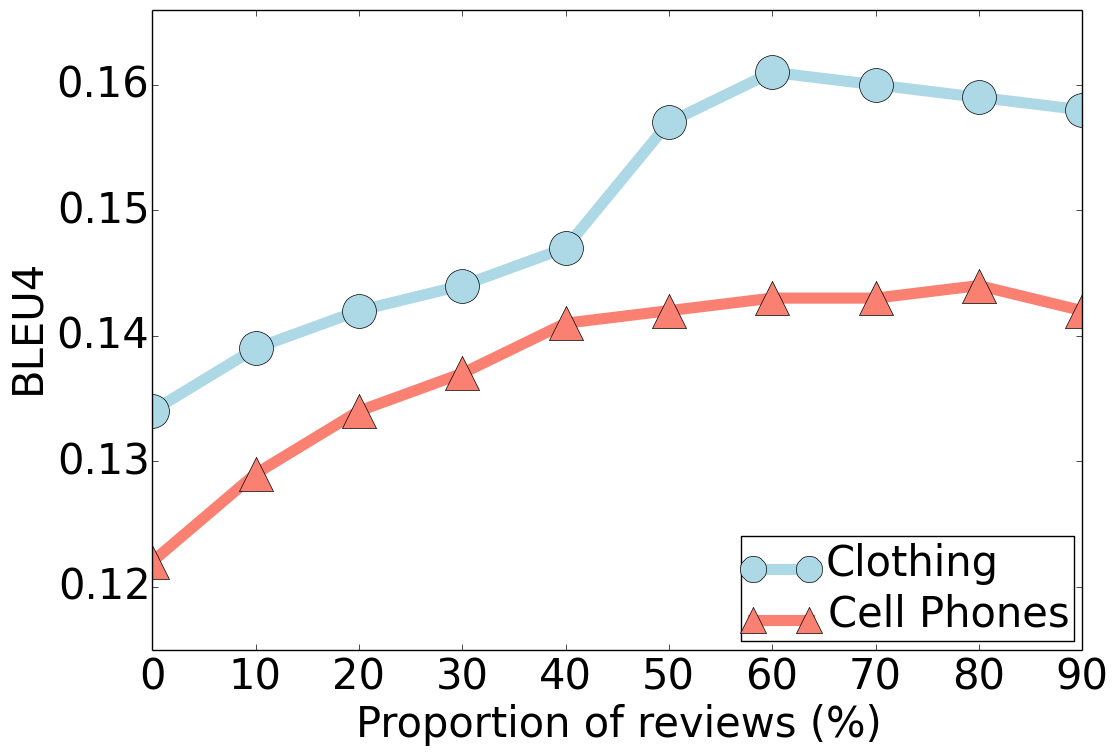}
    \includegraphics[width=0.38\textwidth]{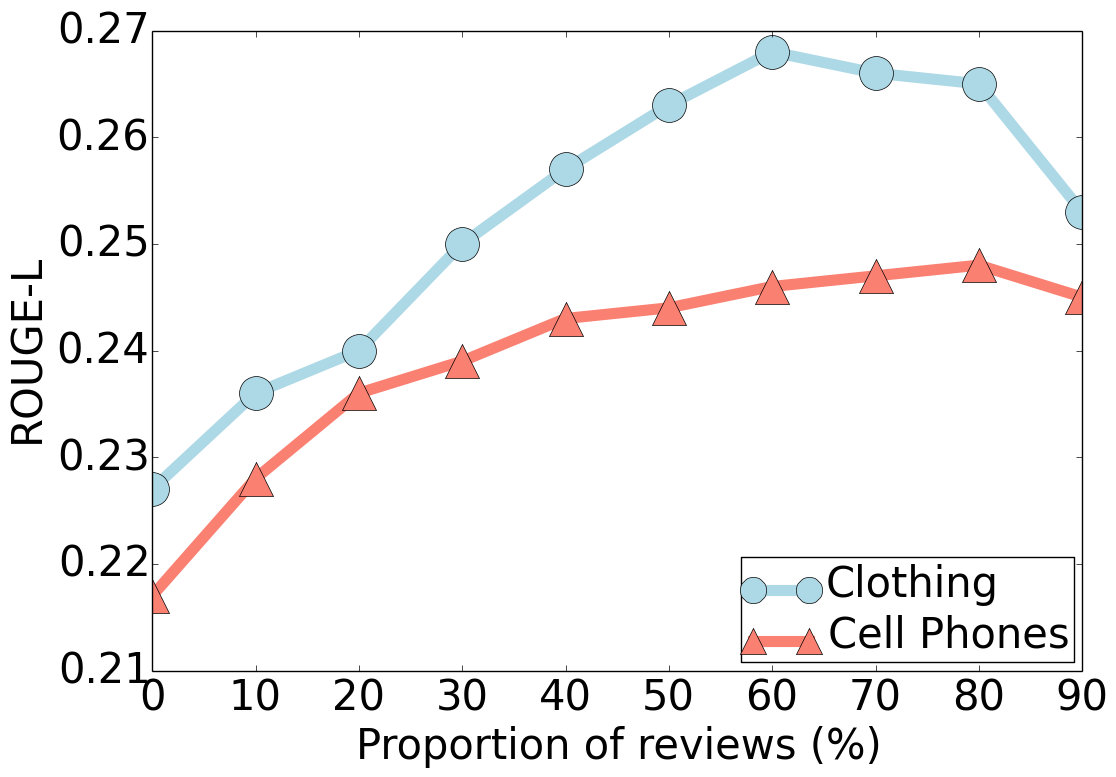}
    \caption{Analysis for proposition of instances.}
    \label{fig:analysis}
\end{figure}

\subsection{Analysis on Instances Composition}
\label{sec:sec:analysis}
The training instance set for the \textit{generator}, i.e., $\textbf{S}$ in Algorithm \ref{alg}, is initialized with QA set and gradually adapted and augmented. Here, we investigate the effect of composition property of $\textbf{S}$ on the \textit{generator} performance at different epochs.
As shown in Fig \ref{fig:analysis}, two product categories and two metrics are illustrated, with the gradually changed training instance set $\textbf{S}$. The proportion of review-question ($qr$) instances in $\textbf{S}$ starts with 0, and significant performance improvement can be observed while the $qr$ proportion gradually increases. The results stay stable until the $qr$ proportion reach 80\%.

\section{Conclusions}
We propose a practical task of question generation from reviews, whose major challenge is the lack of training instances. An adaptive instance transfer and augmentation framework is designed for handling the task via an iterative learning algorithm. Unsupervised aspect extraction is integrated for aspect-aware question generation. Experiments on real-world E-commerce data demonstrate the effectiveness of the training instance manipulation in our framework and the potentials of the review-based question generation task.


\balance
\bibliography{qg-review.bib}

\begin{thebibliography}{38}
\expandafter\ifx\csname natexlab\endcsname\relax\def\natexlab#1{#1}\fi

\bibitem[{Bing et~al.(2016)Bing, Wong, and Lam}]{bing_TOIT_2016}
Lidong Bing, Tak-Lam Wong, and Wai Lam. 2016.
\newblock Unsupervised extraction of popular product attributes from e-commerce
  web sites by considering customer reviews.
\newblock \emph{ACM Transactions on Internet Technology}, 16:1--17.

\bibitem[{Chali and Baghaee(2018)}]{chali2018automatic}
Yllias Chali and Tina Baghaee. 2018.
\newblock Automatic opinion question generation.
\newblock In \emph{INLG}, pages 152--158.

\bibitem[{Chelliah and Sarkar(2017)}]{Chelliah:2017:PRE:3109859.3109936}
Muthusamy Chelliah and Sudeshna Sarkar. 2017.
\newblock Product recommendations enhanced with reviews.
\newblock In \emph{ACM Conference on Recommender Systems}, RecSys '17, pages
  398--399.

\bibitem[{Chen et~al.(2013)Chen, Mukherjee, Liu, Hsu, Castellanos, and
  Ghosh}]{chen2013exploiting}
Zhiyuan Chen, Arjun Mukherjee, Bing Liu, Meichun Hsu, Malu Castellanos, and
  Riddhiman Ghosh. 2013.
\newblock Exploiting domain knowledge in aspect extraction.
\newblock In \emph{EMNLP}, pages 1655--1667.

\bibitem[{Cui et~al.(2017)Cui, Chen, Wei, Wang, Liu, and Hu}]{cui2017attention}
Yiming Cui, Zhipeng Chen, Si~Wei, Shijin Wang, Ting Liu, and Guoping Hu. 2017.
\newblock Attention-over-attention neural networks for reading comprehension.
\newblock In \emph{ACL}, pages 593--602.

\bibitem[{Dong et~al.(2018)Dong, Hong, Chen, Li, Zhang, and
  Zhu}]{dong2018neural}
Xiaozheng Dong, Yu~Hong, Xin Chen, Weikang Li, Min Zhang, and Qiaoming Zhu.
  2018.
\newblock Neural question generation with semantics of question type.
\newblock In \emph{CCF NLPCC}, pages 213--223.

\bibitem[{Du and Cardie(2017)}]{du2017identifying}
Xinya Du and Claire Cardie. 2017.
\newblock Identifying where to focus in reading comprehension for neural
  question generation.
\newblock In \emph{EMNLP}, pages 2067--2073.

\bibitem[{Du and Cardie(2018)}]{du2018harvesting}
Xinya Du and Claire Cardie. 2018.
\newblock Harvesting paragraph-level question-answer pairs from wikipedia.
\newblock In \emph{ACL}, pages 1907--1917.

\bibitem[{Du et~al.(2017)Du, Shao, and Cardie}]{du2017learning}
Xinya Du, Junru Shao, and Claire Cardie. 2017.
\newblock Learning to ask: Neural question generation for reading
  comprehension.
\newblock In \emph{ACL}, pages 1342--1352.

\bibitem[{Duan et~al.(2017)Duan, Tang, Chen, and Zhou}]{duan2017question}
Nan Duan, Duyu Tang, Peng Chen, and Ming Zhou. 2017.
\newblock Question generation for question answering.
\newblock In \emph{EMNLP}, pages 866--874.

\bibitem[{Gao et~al.(2019)Gao, Bing, Chen, Lyu, and King}]{Gao2019DQG}
Yifan Gao, Lidong Bing, Wang Chen, Michael~R. Lyu, and Irwin King. 2019.
\newblock Difficulty controllable generation of reading comprehension
  questions.
\newblock In \emph{IJCAI}, pages 4968--4974.

\bibitem[{Golub et~al.(2017)Golub, Huang, He, and Deng}]{golub2017two}
David Golub, Po-Sen Huang, Xiaodong He, and Li~Deng. 2017.
\newblock Two-stage synthesis networks for transfer learning in machine
  comprehension.
\newblock In \emph{EMNLP}, pages 835--844.

\bibitem[{He et~al.(2017)He, Lee, Ng, and Dahlmeier}]{he2017unsupervised}
Ruidan He, Wee~Sun Lee, Hwee~Tou Ng, and Daniel Dahlmeier. 2017.
\newblock An unsupervised neural attention model for aspect extraction.
\newblock In \emph{ACL}, pages 388--397.

\bibitem[{Hu et~al.(2018)Hu, Liu, Ma, Zhao, and Yan}]{hu2018aspect}
Wenpeng Hu, Bing Liu, Jinwen Ma, Dongyan Zhao, and Rui Yan. 2018.
\newblock Aspect-based question generation.
\newblock In \emph{ICLR Workshop track}.

\bibitem[{Kobayashi(2018)}]{kobayashi2018contextual}
Sosuke Kobayashi. 2018.
\newblock Contextual augmentation: Data augmentation by words with paradigmatic
  relations.
\newblock In \emph{Proceedings of the 2018 Conference of the North American
  Chapter of the Association for Computational Linguistics: Human Language
  Technologies, Volume 2 (Short Papers)}, pages 452--457.

\bibitem[{Lee et~al.(2018)Lee, Heo, and Zhang}]{lee2018answerer}
Sang-Woo Lee, Yu-Jung Heo, and Byoung-Tak Zhang. 2018.
\newblock Answerer in questioner's mind: Information theoretic approach to
  goal-oriented visual dialog.
\newblock In \emph{NeurIPS}, pages 2579--2589.

\bibitem[{Li et~al.(2019)Li, Gao, Bing, King, and Lyu}]{LiGao2019tothepoint}
Jingjing Li, Yifan Gao, Lidong Bing, Irwin King, and Michael~R. Lyu. 2019.
\newblock Improving question generation with to the point context.
\newblock In \emph{EMNLP}, pages 3214--3224.

\bibitem[{Li et~al.(2020)Li, He, Ye, Ng, Bing, and Yan}]{Juntao2020CDCL}
Juntao Li, Ruidan He, Hai Ye, Hwee~Tou Ng, Lidong Bing, and Rui Yan. 2020.
\newblock Unsupervised domain adaptation of a pretrained cross-lingual language
  model.
\newblock In \emph{IJCAI}.

\bibitem[{Li et~al.(2018)Li, Bing, Lam, and Shi}]{li-etal-2018-transformation}
Xin Li, Lidong Bing, Wai Lam, and Bei Shi. 2018.
\newblock Transformation networks for target-oriented sentiment classification.
\newblock In \emph{ACL}, pages 946--956.

\bibitem[{Long et~al.(2017)Long, Zhu, Wang, and Jordan}]{long2017deep}
Mingsheng Long, Han Zhu, Jianmin Wang, and Michael~I Jordan. 2017.
\newblock Deep transfer learning with joint adaptation networks.
\newblock In \emph{ICML}, pages 2208--2217.

\bibitem[{McAuley et~al.(2015)McAuley, Targett, Shi, and Van
  Den~Hengel}]{mcauley2015image}
Julian McAuley, Christopher Targett, Qinfeng Shi, and Anton Van Den~Hengel.
  2015.
\newblock Image-based recommendations on styles and substitutes.
\newblock In \emph{SIGIR}, pages 43--52.

\bibitem[{Pan and Yang(2009)}]{pan2009survey}
Sinno~Jialin Pan and Qiang Yang. 2009.
\newblock A survey on transfer learning.
\newblock \emph{IEEE Transactions on TKDE}, 22(10):1345--1359.

\bibitem[{Pennington et~al.(2014)Pennington, Socher, and
  Manning}]{Pennington2014GloveGV}
Jeffrey Pennington, Richard Socher, and Christopher~D. Manning. 2014.
\newblock Glove: Global vectors for word representation.
\newblock In \emph{EMNLP}, pages 1532--1543.

\bibitem[{Raina et~al.(2007)Raina, Battle, Lee, Packer, and Ng}]{raina2007self}
Rajat Raina, Alexis Battle, Honglak Lee, Benjamin Packer, and Andrew~Y Ng.
  2007.
\newblock Self-taught learning: transfer learning from unlabeled data.
\newblock In \emph{ICML}, pages 759--766. ACM.

\bibitem[{Sun et~al.(2018)Sun, Liu, Lyu, He, Ma, and Wang}]{sun2018answer}
Xingwu Sun, Jing Liu, Yajuan Lyu, Wei He, Yanjun Ma, and Shi Wang. 2018.
\newblock Answer-focused and position-aware neural question generation.
\newblock In \emph{EMNLP}, pages 3930--3939.

\bibitem[{Tan et~al.(2017)Tan, Zhang, Pan, and Yang}]{tan2017distant}
Ben Tan, Yu~Zhang, Sinno~Jialin Pan, and Qiang Yang. 2017.
\newblock Distant domain transfer learning.
\newblock In \emph{AAAI}, pages 2604--2610.

\bibitem[{Tang et~al.(2017)Tang, Duan, Qin, Yan, and Zhou}]{tang2017question}
Duyu Tang, Nan Duan, Tao Qin, Zhao Yan, and Ming Zhou. 2017.
\newblock Question answering and question generation as dual tasks.
\newblock \emph{arXiv preprint arXiv:1706.02027}.

\bibitem[{Tang et~al.(2018)Tang, Duan, Yan, Zhang, Sun, Liu, Lv, and
  Zhou}]{tang2018learning}
Duyu Tang, Nan Duan, Zhao Yan, Zhirui Zhang, Yibo Sun, Shujie Liu, Yuanhua Lv,
  and Ming Zhou. 2018.
\newblock Learning to collaborate for question answering and asking.
\newblock In \emph{NAACL-HLT}, pages 1564--1574.

\bibitem[{Wan and McAuley(2016)}]{wan2016modeling}
Mengting Wan and Julian McAuley. 2016.
\newblock Modeling ambiguity, subjectivity, and diverging viewpoints in opinion
  question answering systems.
\newblock In \emph{ICDM}, pages 489--498.

\bibitem[{Wang et~al.(2015)Wang, Liu, Cao, Zhao, and
  De~Melo}]{wang2015sentiment}
Linlin Wang, Kang Liu, Zhu Cao, Jun Zhao, and Gerard De~Melo. 2015.
\newblock Sentiment-aspect extraction based on restricted boltzmann machines.
\newblock In \emph{ACL}, pages 616--625.

\bibitem[{Wang et~al.(2019)Wang, Wei, Fan, Liu, and Huang}]{wang2019a}
Siyuan Wang, Zhongyu Wei, Zihao Fan, Yang Liu, and Xuanjing Huang. 2019.
\newblock A multi-agent communication framework for question-worthy phrase
  extraction and question generation.
\newblock In \emph{AAAI}, pages 7168--7175.

\bibitem[{Wang et~al.(2018)Wang, Lan, Nie, Waters, Grimaldi, and
  Baraniuk}]{wang2018qg}
Zichao Wang, Andrew~S Lan, Weili Nie, Andrew~E Waters, Phillip~J Grimaldi, and
  Richard~G Baraniuk. 2018.
\newblock {QG-Net}: a data-driven question generation model for educational
  content.
\newblock In \emph{Annual ACM Conference on Learning at Scale}, page~7.

\bibitem[{Xiao et~al.(2018)Xiao, Wang, Feng, and Zheng}]{xiao2018dual}
Han Xiao, Feng Wang, Yanjian Feng, and Jingyao Zheng. 2018.
\newblock Dual ask-answer network for machine reading comprehension.
\newblock \emph{arXiv preprint arXiv:1809.01997}.

\bibitem[{Yang et~al.(2017)Yang, Hu, Salakhutdinov, and Cohen}]{yang2017semi}
Zhilin Yang, Junjie Hu, Ruslan Salakhutdinov, and William Cohen. 2017.
\newblock Semi-supervised qa with generative domain-adaptive nets.
\newblock In \emph{ACL}, pages 1040--1050.

\bibitem[{Yu et~al.(2018)Yu, Dohan, Luong, Zhao, Chen, Norouzi, and
  Le}]{yu2018qanet}
Adams~Wei Yu, David Dohan, Minh-Thang Luong, Rui Zhao, Kai Chen, Mohammad
  Norouzi, and Quoc~V Le. 2018.
\newblock {QANet}: Combining local convolution with global self-attention for
  reading comprehension.
\newblock In \emph{ICLR}.

\bibitem[{Zhao et~al.(2018{\natexlab{a}})Zhao, Guan, Chen, He, Cai, Wang, and
  Wang}]{zhao2018weakly}
Wei Zhao, Ziyu Guan, Long Chen, Xiaofei He, Deng Cai, Beidou Wang, and Quan
  Wang. 2018{\natexlab{a}}.
\newblock Weakly-supervised deep embedding for product review sentiment
  analysis.
\newblock \emph{IEEE Transactions on TKDE}, 30(1):185--197.

\bibitem[{Zhao et~al.(2018{\natexlab{b}})Zhao, Ni, Ding, and
  Ke}]{zhao2018paragraph}
Yao Zhao, Xiaochuan Ni, Yuanyuan Ding, and Qifa Ke. 2018{\natexlab{b}}.
\newblock Paragraph-level neural question generation with maxout pointer and
  gated self-attention networks.
\newblock In \emph{EMNLP}, pages 3901--3910.

\bibitem[{Zhou et~al.(2017)Zhou, Yang, Wei, Tan, Bao, and
  Zhou}]{zhou2017neural}
Qingyu Zhou, Nan Yang, Furu Wei, Chuanqi Tan, Hangbo Bao, and Ming Zhou. 2017.
\newblock Neural question generation from text: A preliminary study.
\newblock In \emph{CCF NLPCC}, pages 662--671.

\end{thebibliography}
\bibliographystyle{acl_natbib}

\end{document}